\documentclass[twocolumn,showpacs,preprintnumbers,amsmath,amssymb]{revtex4}
\usepackage{graphicx}
\usepackage{dcolumn}
\usepackage{bm}

\usepackage{color}

\begin{document}

\title{On the inverse cascade of magnetic helicity}
\author{Alexandros Alexakis} \email{alexakis@ucar.edu}
\author{Pablo Mininni}       \email{mininni@ucar.edu}
\author{Annick Pouquet}      \email{pouquet@ucar.edu}
\affiliation{National Center for Atmospheric Research}
\date{\today}

\begin{abstract}
We study the inverse cascade of magnetic helicity in conducting fluids 
by investigating the detailed transfer of helicity between different 
spherical shells in Fourier space in direct numerical simulations 
of three-dimensional magnetohydrodynamics (MHD). Two 
different numerical simulations are used, one where the system is forced 
with an electromotive force in the induction equation, and one in which the
system is forced mechanically with an ABC flow and the magnetic field is 
solely sustained by a dynamo action. The magnetic helicity cascade 
at the initial stages of both simulations is observed to be inverse 
and local (in scale space) in the large scales, and direct and local in 
the small scales. When saturation is approached most of the helicity 
is concentrated in the large scales and the cascade is non-local. Helicity is 
transfered directly from the forced scales to the largest scales. At the 
same time, a smaller in amplitude direct cascade is observed from 
the largest scale to small scales. 
\end{abstract}

\maketitle

\newpage
\section{ \label{Intro} Introduction }
The generation of magnetic fields in various astrophysical objects ranging 
from planets (e.g. the geodynamo \cite{Glatzmaier96,Kono02}), to stars 
(e.g the solar dynamo \cite{Dikpati99,Nandy02,Bushby04}), and spiral 
galaxies (e.g. the interstellar dynamo \cite{Shukurov05}), is mostly 
attributed to dynamo action due to motions of a conducting fluid 
\cite{Moffatt}. 
Due to the magnetic flux conservation in ideal magnetohydrodynamics 
(MHD), the stretching of magnetic field lines by a conducting flow 
amplifies the magnetic energy at small scales. To further explain how 
magnetic fields end up in scales much larger than the outer scales 
of fluid motions, one of the theoretical arguments used is the 
inverse cascade of magnetic helicity in MHD turbulence. It is worth mentioning 
here that the the presence of helicity in the flow, although helpful, 
is not required to generate large scale magnetic fields. In some 
circumstances, large scale fields can be sustained solely by helicity fluctuations 
\citep{Gilbert88}, by anisotropic flows \citep{Nore97}, or large scale shear \citep{Urpin02}.

Early studies using mean-field theory \cite{Steenbeck66,Krause} 
and turbulent closure models \cite{Frish75,Pouquet76} have shown within 
the framework of the approximations made that magnetic helicity cascades 
inversely from small scales to large scales. Direct numerical simulations 
(DNS) \cite{Pouquet78,Meneguzzi81,Kida91,Brandenburg01,Gomez04} have 
verified the inverse cascade of magnetic helicity and have shown the 
generation of large scale magnetic fields from small scale helical forcing. 
A detailed examination of the cascading process was investigated in 
\cite{Brandenburg01}, where the rate of transfer of magnetic energy among 
different scales was measured from DNS. The results showed evidence 
of nonlocal energy transfer of magnetic energy from the small scales to 
the large scales, suggesting also a nonlocal transfer of magnetic 
helicity. However, in three dimensional MHD turbulence the ideal 
invariant that can display an inverse cascade, {\it stricto sensu},
is the magnetic helicity, 
not the magnetic energy, and no direct attempt to measure its transfer 
in simulations has been done so far.

In this paper we focus on helical flows, and revisit the 
problem of the inverse cascade of the magnetic helicity by analyzing 
two DNS: one forced through the induction equation by an external 
electromotive force, and one forced mechanically. 
In both simulations, the forcing was applied in small scales so that enough 
large scales were available for an inverse cascade to develop. 
Note that this election naturally limits the 
Reynolds numbers we can resolve, and as a result only moderate Reynolds 
numbers will be considered in this work. Extending the formalism used 
in \cite{Dar01,Debliquy05,Alexakis05,Mininni05} 
for the transfer of the magnetic and kinetic energy, 
we directly measured the transfer rate of magnetic helicity among different scales, both 
in scales larger and smaller than the forcing scales. 

The outline of the paper is as follows. In Sec. \ref{Theory} we 
present a brief review of the equations and definition of transfer 
functions needed to study this problem. In Sec. \ref{Magnetically_Forced} 
we give the results from the magnetically forced simulation; and in Sec. 
\ref{Mechanically_Forced} we give the results from the mechanically 
forced simulation. Finally, we discuss the implications of our 
results in Sec. \ref{Concs}, where we also give our conclusions.

\section{ \label{Theory} Theory and definitions}

To a good approximation the equations that describe the dynamics of an 
incompressible conducting fluid coupled to a magnetic field are given by:
\begin{equation}
\label{NS}
\partial_t {\bf u} + {\bf u}\cdot \nabla {\bf u} = 
- \nabla P + {\bf b}\cdot \nabla {\bf b}
+ \nu \nabla^2 {\bf u} +{\bf f}
\end{equation}
\begin{equation}
\partial_t {\bf b} = \nabla \times ({\bf u} \times {\bf b})
+ \eta \nabla^2 {\bf b} + \nabla \times {\bf E}
\label{magn}
\end{equation}
where ${\bf u}$ is the velocity field, ${\bf b}$ is the magnetic field, 
$\nu$ is the kinematic viscosity, $\eta$ is the magnetic diffusivity, 
$P$ is the total pressure, ${\bf f}$ an external mechanic force, and 
${\bf E}$ is an external electromotive force. The equations are 
written in the familiar Alfv\'enic dimensionless units. These equations 
are accompanied by the conditions
$\nabla \cdot {\bf u} = 0 = \nabla \cdot {\bf b}$. 
This last condition allows us to write the magnetic field in terms of 
a vector potential ${\bf b= \nabla \times a}$. Removing a curl from 
Eq. (\ref{magn}), the evolution equation for the vector potential reads:
\begin{equation}
\partial_t {\bf a} = {\bf u} \times {\bf b} 
+ \eta \nabla^2{\bf a} - \nabla \phi +{\bf E}
\label{vec_p}
\end{equation}
where the Coulomb gauge ($\nabla \cdot {\bf a}=0$) is assumed
and $\nabla \phi$ is determined by the solenoidal condition on ${\bf a}$.
There are three quadratic invariants in the absence of dissipation
and forcing: the total energy $E=\int ({\bf b^2+u^2})/2 \, dx^3$, 
the cross-helicity $H_c=\int {\bf b\cdot u} \, dx^3$, and the magnetic 
helicity $H_m=\int {\bf b \cdot a}/2 \, dx^3$. To the best 
of our knowledge the magnetic helicity, which is the quantity under 
investigation in this paper, was first introduced as an invariant of the MHD equations by Woltjer 
\cite{Woltjer58}. It is proportional to the number of linkages of the 
magnetic field lines \cite{Moffatt}, as is reflected by its relation 
with topological quantities as the Gauss linking number 
\cite{Wright89,Berger97}. The conservation of magnetic helicity is 
related with the frozen-in theorem of Alfv\'en. Being magnetic field lines 
material, a link can only change through reconnection of field lines, 
and therefore breaking of the frozen-in condition ({\it e.g.} through  dissipation) is 
needed.

As we stated in the introduction, we want to quantify the rate at which
helicity is transfered among the different scales of the magnetic field. 
To define the magnetic field and vector potential at different scales, 
we introduce the shell-filtered magnetic field and vector potential 
components ${\bf b}_K({\bf x})$ and ${\bf a}_K({\bf x})$. Here, the 
subscript $K$ indicates that the field has been filtered to keep 
only the modes whose wave vectors are in the Fourier shell $[K,K+1]$ 
(hereafter called the shell $K$). Clearly the sum of all the $K$ components 
gives back the original field, ${\bf b}=\sum_K {\bf b}_K$, and the 
filtering operation commutes with the curl operator, 
$\nabla \times {\bf a}_K =  {\bf b}_K$. A similar decomposition has 
been used in \cite{Alexakis05,Mininni05} to study the cascade of energy.

We are interested  in the rate that magnetic helicity at a given 
shell $Q$ is transferred into a different shell $K$. From the MHD 
equations, taking the dot product of Eq. (\ref{vec_p}) with 
${\bf b}_K/2$, taking the dot product of Eq. (\ref{magn}) with 
${\bf a}_K/2$, adding them, and integrating over space, we finally 
obtain the evolution of the magnetic helicity 
$H_m(K)=\frac{1}{2}\int {\bf a}_K\cdot {\bf b}_K \, dx^3$ in the shell $K$:
{\setlength\arraycolsep{2pt}
\begin{eqnarray}
&& \partial_t H_{m}(K) = 
    \sum_Q \int {\bf b}_K \cdot ({\bf u \times b}_Q) \, d{\bf x}^3 
- {} \nonumber \\
&&{} - \eta \int {\bf b}_K \cdot {\bf \nabla \times b}_K \, d{\bf x}^3 
    + \int {\bf b}_K \cdot {\bf E}_K \, d{\bf x}^3 \,\, .
\label{EbI}
\end{eqnarray}}
We can rewrite Eq. (\ref{EbI}) in a more compact form
\begin{equation}
\partial_t { H}_{m}(K) = 
\sum_Q {\mathcal T}_{h}(K,Q) - 
\eta {\mathcal D}_h(K) + {\mathcal F}_h(K) ,
\label{eq:Eb}
\end{equation}
where we have introduced the transfer function ${\mathcal T}_{h}(Q,K)$, 
the helicity injection ${\mathcal F}_h(K)$, and  
the helicity dissipation ${\mathcal D}_h(K)$ 
as defined below.

The dissipation of magnetic helicity in the shell $K$ is given by
\begin{equation}
{\mathcal D}_h(K) = \int{ {\bf b}_K \cdot (\nabla \times {\bf b}_K) 
    d{\bf x}^3} ,
\end{equation}
Note however that unlike the energy dissipation this is not a positive
definite quantity.

The injection rate of magnetic helicity in the shell $K$ by the 
external electromotive force ${\bf E}$ is given by
\begin{equation}
{\mathcal F}_h(K) = \int {\bf b}_K \cdot {\bf E}_K \, d{\bf x}^3 .
\label{eq:Fh}
\end{equation}
Note that the mechanical forcing ${\bf f}$ does not inject magnetic 
helicity in the system, as follows from Eq. (\ref{EbI}). However, 
as will be discussed later, if the external mechanical forcing is helical, 
the velocity field can generate helical magnetic fields locally through 
the ${\mathcal T}_{h}(Q,K)$ term.

The transfer rate of magnetic helicity at shell $Q$ into 
magnetic helicity at shell $K$ is defined as:
\begin{equation}
{\mathcal T}_{h}(K,Q) = \int{ {\bf b}_K \cdot ({\bf u} \times {\bf b}_Q)
    d{\bf x}^3} . \label{eq:Th}
\end{equation}
${\mathcal T}_{h}(K,Q)$ expresses the transfer rate of positive helicity  
from the shell $Q$ to the shell $K$, or equivalently the transfer rate 
of negative helicity from the shell $K$ into the shell $Q$. Positive 
values of ${\mathcal T}_{h}(K,Q)$ imply that positive helicity is 
transfered from the shell $Q$ to the shell $K$, while negative values 
imply the reverse transfer. The transfer term is a conservative term 
and it does not generate or destroy total magnetic helicity. However, 
this term is responsible for the redistribution of magnetic helicity 
among different scales. This fact is expressed by the anti-symmetry 
property of ${\mathcal T}_{h}(Q,K)$:
\begin{equation}
 {\mathcal T}_{h}(K,Q)=- {\mathcal T}_{h}(Q,K) \ .
\end{equation} 
We stress that helicity (unlike energy) is not a 
positive definite quantity and care needs to be taken when we interpret 
results. We will not attempt here a separation of its different sign 
components (see e.g. \cite{Waleffe91,Chen03a,Chen03b} for the kinetic 
helicity in hydrodynamic turbulence).
As an example, if in some shell $K$ the helicity is positive 
$H_b(K)>0$ with a positive rate of change $\partial_t { H}_{b}(K)>0$, 
then the magnetic field becomes more helical in that shell as the 
system evolves. If however the helicity is negative $H_b(K)<0$, then 
positive rate of change implies that the field becomes less helical 
in that shell. 
In the same spirit, if positive helicity is transfered from scales with 
negative helicity to scales with positive helicity, the field becomes more 
helical at both scales even if the total helicity at all scales remains 
constant. On the other hand, if positive helicity is transfered from 
scales with positive helicity to scales with negative helicity, the field 
becomes less helical at each scale since the absolute value of magnetic 
helicity in each scale is decreased.


\section{ \label{Magnetically_Forced} Magnetically Forced Run}

We begin with the magnetically forced simulations. In this case, a helical ABC electromotive force 
${\bf E}$ is used, while keeping the mechanical force equal to zero. The 
flow evolution is solved using a pseudospectral method with the 
$2/3$-rule for dealiasing on a $N^3 = 256^3$ grid. No uniform 
magnetic fields is allowed in the periodic box, and therefore magnetic 
helicity conservation is satisfied in the ideal case \cite{Berger97}.

The viscosity and diffusivity are set to 
$\nu = \eta = 5 \times 10^{-4}$. Only wavenumbers in the shells 8 and 
9 are forced. The phase of the external ABC electromotive force 
is changed randomly with a correlation time $\tau =  1.25 \times 10^{-2}$, 
and the time step to satisfy the Courant-Friedrichs-Levy (CFL) condition 
is set to $\Delta t = 2.5 \times 10^{-3}$. The integral lenghtscale of 
the flow $L = 2 \pi \int E(K)/K dk /E$ (where $E(K)$ is the kinetic energy in 
the shell $K$, and $E$ is the total kinetic energy) in the steady state of the 
simulation is $L \approx 1.42$, and the large scale turnover time 
$T = U/L$ (where $U$ is the r.m.s. velocity) is $T \approx 0.25$. 
Based on these numbers, the resulting kinetic Reynolds number $R_e$ and 
magnetic Reynolds number $R_m$ are $R_e \approx R_m \approx 700$. 
The simulation is well resolved, in the sense that the Kolmogorov's 
kinetic [$k_\nu = (\epsilon/\nu^3)^{1/4}$, where $\epsilon$ is the total 
energy injection rate] and magnetic [$k_\eta = (\epsilon/\nu^3)^{1/4}$] 
dissipation wavenumbers are smaller than the maximum resolved wavenumber 
$k_{max} \approx N/3$.

\begin{figure}
\includegraphics[width=8cm]{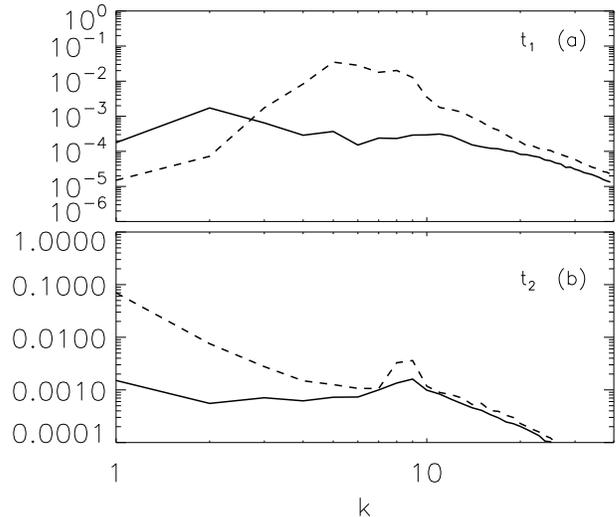}
\caption{\label{fig_01} The kinetic (solid line) and magnetic (dashed line)
energy spectra for the magnetically forced simulation and for two different 
times: panel (a) $t_1$ before saturation is reached, panel (b) $t_2$ 
close to saturation. The figure shows the spectrum up to $k=40$, note 
however that the maximum wavenumber resolved in the code is 
$k_{max} \approx 85$.}
\end{figure}

The magnetically forced case is easier to analyze because only 
one sign of helicity appears to dominate all scales. In Fig. \ref{fig_01} 
we show the resulting spectra at two different times $t_1$ and $t_2$, up to 
wavenumber $k=40$. The former time is early (before the system 
comes close to saturation), and the latter time is when the flow is almost 
saturated. Note that the maximum wavenumber in the code is 
$k_{max} \approx 85$, and the dissipative range in the simulation extends 
to larger wavenumbers than what is shown in Fig. \ref{fig_01}. However, 
the transfer function ${\mathcal T}_h(K,Q)$ will only be computed up to 
$K,Q = 40$, a range that includes all scales larger (wavenumbers smaller) 
than the injection band, as well as the scales smaller than the injection 
band where a turbulent scaling can still be identified. We will follow 
this convention in the following figures.

\begin{figure}
\includegraphics[width=8cm]{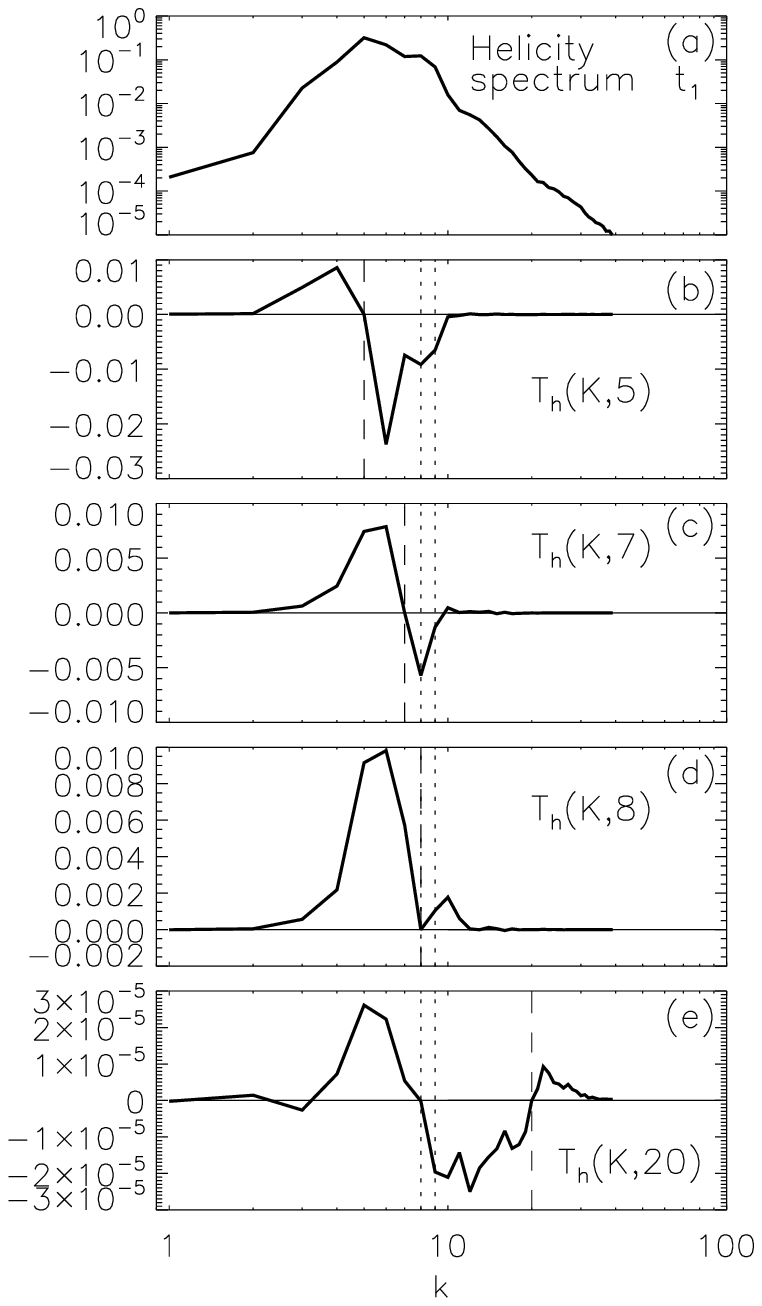}
\caption{\label{fig_02} The magnetic helicity (panel a) and its transfer
 ${\mathcal T}_h(K,Q)$  from shell $Q$ to shell $K$,normalized by the total magnetic 
helicity for the magnetically forced run at time $t_1$. The different panels (b-e) 
correspond to different values of $Q=5,7,8,20$.
The dashed vertical line indicates the location of the examined value of $Q$
while the dotted lines give the width of the forcing band. 
Note that the transfer for $Q=20$ is significantly smaller.}
\end{figure}

\subsection{Early times}

The spectrum of magnetic helicity for $t=t_1$ is shown in Fig. 
\ref{fig_02}(a), in log-log scale and is 
positive at all scales. At this stage, the magnetically helicity spectrum 
peaks at wave number $K\simeq5$, a scale slightly larger than the scale 
where the system is forced. Panels (b-e) show the transfer of magnetic 
helicity ${\mathcal T}_h(K,Q)$ at different values of $Q$, 
normalized by the total magnetic helicity in that shell. The dotted 
lines in these panels indicate the shell where the forcing is applied, 
while the dashed lines indicate the mode that is examined.

Since the helicity is positive for all scales, we only need to interpret 
${\mathcal T}_h$ as transfer of positive helicity. In Fig. \ref{fig_02}, 
panel (b) shows the transfer ${\mathcal T}_h(K,Q)$ for wavenumbers at the peak of the energy 
spectrum ($K=5$). For smaller wavenumbers ($K<5$) the transfer is positive, 
while it is negative for larger wave numbers ($K>5$). This picture indicates 
that the shell $K=5$ is giving/transferring helicity to its close neighbors 
on the left, while it receives helicity from its neighbors on the right. 

Similar behavior is observed for the modes with wavenumbers between the 
peak of magnetic helicity in Fourier space and the forcing wave 
number [see panel (a)]. The transfer of magnetic helicity 
for a value of $Q$ in this range ($Q=7$) is shown in panel (c), suggesting 
the picture of a local inverse cascade. Indeed, the shell $Q=7$ gives 
most of its helicity to the shell $K \approx 6$ (positive peak), while 
receives helicity from the shell $K \approx 8$ (negative peak).

The forced wave numbers [the transfer ${\mathcal T}_h(K,Q)$ for 
$Q=8$ is shown in panel (d)] are giving helicity to both smaller and 
larger scales, with a preference towards the larger scales (smaller 
wave numbers). Finally, wavenumbers larger than the forced scale 
[panel (e)] have a different behavior. Unlike the large scales, the small 
scales ($K=20$ is displayed here) receive helicity from larger scales 
(but smaller than the forced scale) and give helicity to smaller scales. 
This suggests a local direct cascade of positive magnetic helicity. 
In addition, there is a non-local transfer of helicity to much larger 
scales ($K\simeq5$), probably associated to reconnection events.   

\begin{figure}
\includegraphics[width=8cm]{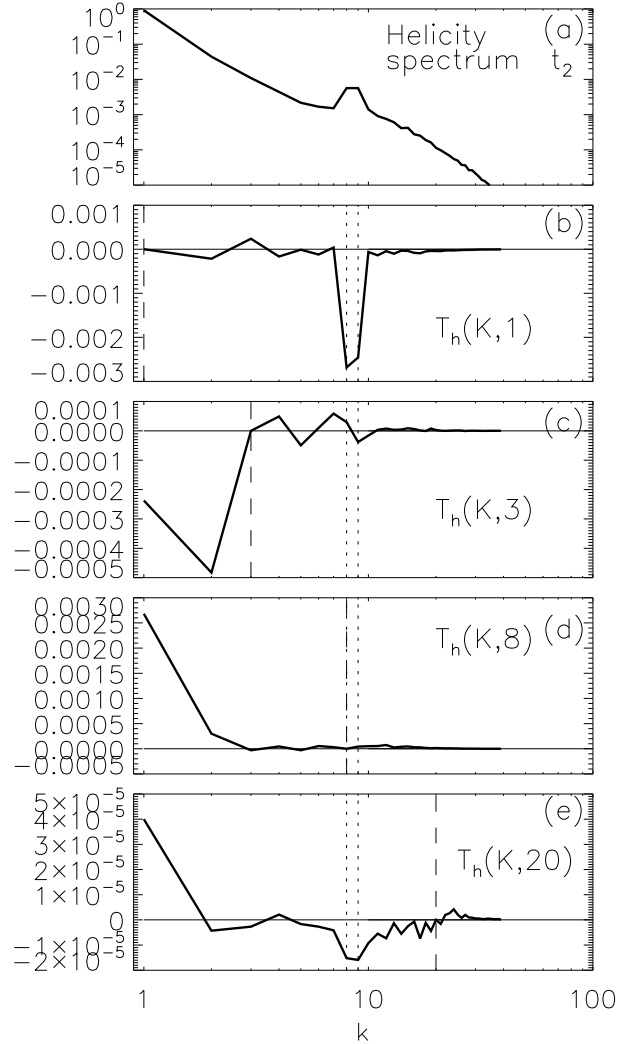}
\caption{\label{fig_03} The magnetic helicity (panel a)
and its transfer $T_h(K,Q)$ from shell $Q$ to shell $K$, normalized by the total
magnetic helicity for the magnetically forced run, at later times.
The different panels (b-e) correspond to different values of $Q=1,3,8,20$
The dashed vertical line indicates the location of the examined value of $Q$
while the dotted lines give the width of the forcing band. 
Note the different values on the vertical axis.}
\end{figure}

\subsection{Late times}

The picture of local cascade of helicity is changed at later times, as 
the peak of the helicity spectrum moves to the largest possible scale 
($K=1$). The helicity spectrum and the transfer functions at this stage 
are shown in Fig. \ref{fig_03}. The helicity spectrum peaks strongly at 
$K=1$ [see panel (a)]. As shown in panel (b), the largest scales are now 
receiving magnetic helicity directly from the forced scales, the 
remaining scales giving almost zero contribution. This behavior suggests a 
non-local inverse cascade. Intermediate scales between the largest 
available and the forced scales have also changed their behavior [panel (c)]. 
These scales ($1<K<8$) now are receiving positive helicity mostly from the 
largest modes and giving it to the smaller scales. This suggests a 
direct cascade of positive magnetic helicity in the range $1<K<8$. It 
appears therefore that once the magnetic helicity has reached the 
largest possible scale there may be some ``reflection" at $K=1$, and helicity 
then cascades to smaller scales with the exception of the forced modes 
that continue feeding the magnetic helicity at the largest 
scale [see panel (d)]. Scales smaller than the injection band transfer 
the positive magnetic helicity to smaller scales as at early 
times, with the exception of a non-local direct input to the largest 
scale.

The ``reflection" of magnetic helicity in Fourier space when it 
reaches the largest scale in the box, suggests that the late time 
evolution is strongly dependent on the boundary conditions. In our case, 
the periodic boundary conditions do not allow magnetic helicity to grow 
at scales even larger, and forbids the system to eject magnetic helicity 
outside the box. Similar behavior has been observed in two-dimensional 
hydrodynamic turbulence, where the quantity which has an inverse cascade 
is the energy \cite{Smith94,Borue94,Boffetta00}. For this latter problem, 
evidence of nonlocal and irregular transfer of the inverse cascading 
invariant was also found in simulations \cite{Danilov01}. We will 
come back to this issue later.

\section{ \label{Mechanically_Forced} Mechanically Forced Run}

We move next to the case where the system is mechanically forced, and 
the magnetic field is solely amplified and sustained against 
Ohmic dissipation by dynamo action. This case is more relevant to most 
physical situations. In this case, we perform a numerical simulation 
using a grid of $N^3=256^3$ points under the following procedure. First, 
a hydrodynamic simulation was performed mechanically forcing at wave 
number $K=10$, with an ABC flow to obtain a turbulent steady state. 
The kinetic helicity of the flow 
$H_k = \int {\bf u} \cdot \nabla \times {\bf u}/2 \, d^3x$ in the steady 
state is positive, and close to maximal. Unlike the previous section, 
here the phases of the ABC flow are kept constant as a function of time.

After reaching the hydrodynamic steady state, a random, 
non-helical, and small magnetic field was introduced and the 
simulation was carried keeping the force fixed to observe the 
evolution of the system from the initial exponential amplification of 
the magnetic energy until the large-scale saturation. The 
kinematic viscosity and magnetic diffusivity were 
$\eta = \nu = 2.5 \times 10^{-3}$. In the hydrodynamic steady state, 
the integral scale of the flow was $L \approx 0.6$ and the large scale 
eddy turnover time $T \approx 0.6$. Based on these numbers, the mechanic 
and magnetic Reynolds numbers were $R_e = R_m = 240$.

From Eqs.
(\ref{eq:Eb}) and (\ref{eq:Fh}), we note that a helical mechanical force
cannot inject net magnetic helicity in the system. However, a flow with positive
kinetic helicity in the forcing band generates equal amounts of magnetic
helicity at large and small scales with opposite signs.
This generation can be understood in a geometrical way from
the Stretch Twist -Fold (STF) dynamo \cite{Zeldovich}. As magnetic flux tubes at large scales
are twisted in one direction (generating one sign of magnetic helicity),
magnetic field lines at small scales are twisted in the opposite
direction.

This generation of opposite signs of magnetic helicity at different
scales is also a signature of the $\alpha$-effect \cite{Seehafer96}. 
In mean field theory (see{\it  e.g} \cite{Krause}) the equation for the evolution 
of the mean magnetic helicity $\overline{H}_m$ is 
\begin{equation}
\partial_t \overline{H}_m = \int \left(\alpha \overline{B}^2 - 
    \beta \overline{\bf B} \cdot \nabla \times \overline{\bf B}\right) 
    d^3 x .
\end{equation}
where $\alpha \approx -\tau \left<{\bf v}\cdot\nabla\times{\bf v}\right>$
is proportional to minus the kinetic helicity of the flow (here ${\bf v}$
is the fluctuating velocity field, and $\tau$ is a correlation time). The
coefficient $\beta$ is a positive turbulent diffusivity. 
As a result, the $\alpha$-effect injects magnetic helicity of opposite
sign than the kinetic helicity into the mean (large scale) magnetic field. As its
counterpart, at small scales the fluctuating magnetic field receives
magnetic helicity of the same sign than the kinetic helicity.

We will investigate three different times. In the first case, the dynamo is 
still kinematic ({\it i.e.} the magnetic energy is smaller than the kinetic 
energy at all scales, and the effect of the Lorentz force on the 
velocity field can thus be neglected). In the second, kinetic and magnetic 
energies are of the same order but the peak of the magnetic helicity is 
not at the largest scales yet. In this regime, scales smaller than 
the energy injection band have reached saturation, while the large scale 
magnetic field keeps growing slowly. Finally, we investigate the 
saturated stage where the magnetic helicity spectrum peaks at the 
largest attainable scale.

\begin{figure}
\includegraphics[width=8cm]{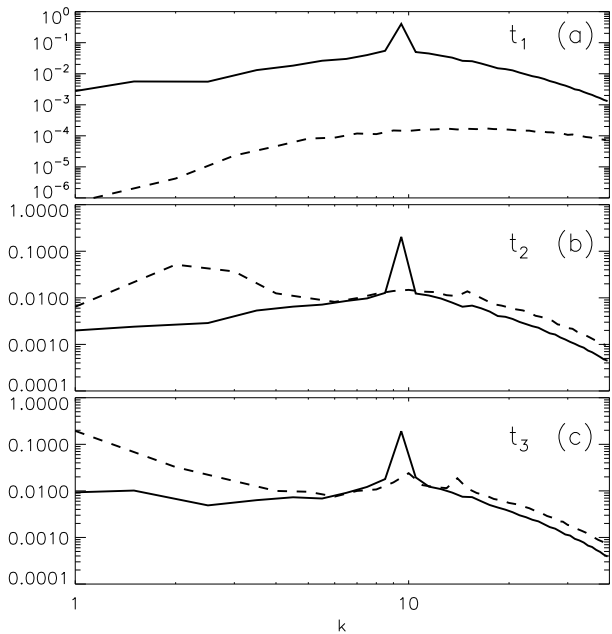}
\caption{\label{fig_04} The kinetic energy spectrum (solid line)
and the magnetic energy spectrum (dashed line) for three different times 
for the mechanically forced run. Spectra are shown up to $k=40$, 
the maximum wavenumber for which the transfer was analyzed. The maximum 
wavenumber resolved in the simulation was $k_{max} \approx 85$.}
\end{figure}

The energy spectra for these three cases are shown in Fig. \ref{fig_04}. 
As in the previous section, the maximum wavenumber resolved in the 
simulation was $k_{max} \approx 85$, and at all times the Kolmogorov's 
dissipation wavenumbers were smaller than $k_{max}$. Since the transfer 
between different shells will only be studied up to $K,Q = 40$, all 
spectral quantities in the figures are shown up to this wavenumber.

\subsection{Kinematic regime}
\begin{figure}
\includegraphics[width=8cm]{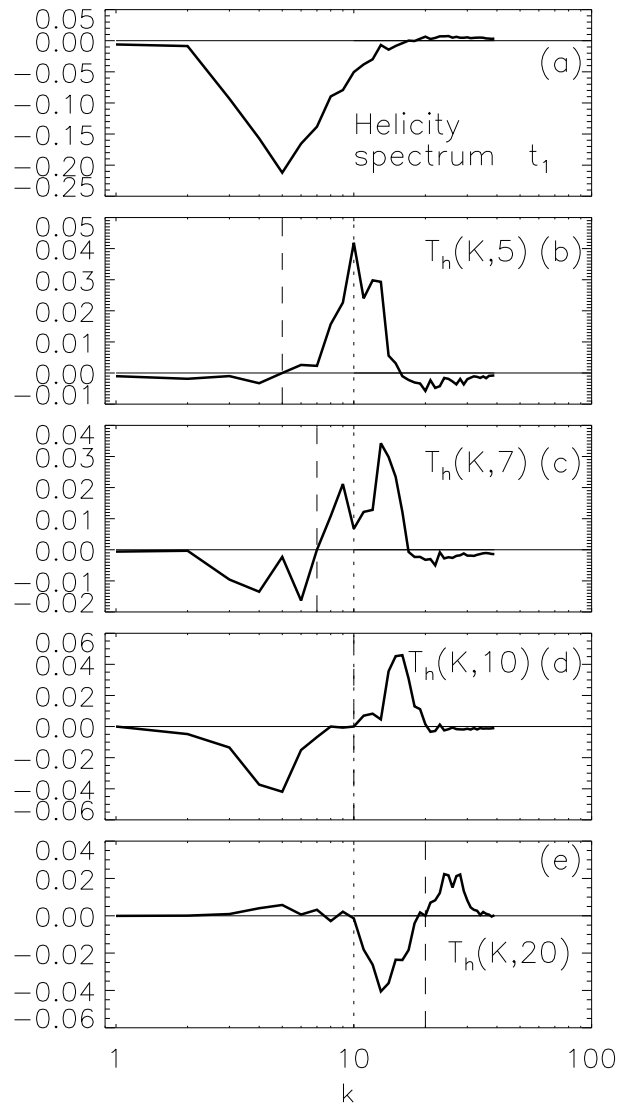}
\caption{\label{fig_05} The spectrum of magnetic helicity (panel a), 
and transfer of magnetic helicity ${\mathcal T}_h(K,Q)$ normalized by 
the total magnetic helicity at the shell $Q$ for the mechanically 
forced run in the kinematic regime. The different panels (b-e) correspond 
to different values of $Q=5,7,10,20$. The dashed vertical line indicates 
the location of the examined value of $Q$ while the dotted line indicates 
the shell where the system was forced. }
\end{figure}

We begin with the kinematic regime. The magnetic helicity spectrum 
is shown in Fig. \ref{fig_05}(a). Unlike the magnetically forced case, 
the magnetic helicity spectrum changes sign. For scales smaller 
than the forced scales, the magnetic helicity spectrum is positive, 
while at large scales the magnetic helicity is negative. The 
positive and negative peaks are close on either side of the forced band. 
The transfer of helicity ${\mathcal T}_h(K,Q)$ for various shells $Q$ is 
shown in Figs. \ref{fig_05}(b-e).

The large scales [$Q=5$ is shown in panel (b)], where the negative 
peak of the magnetic helicity spectrum is located, receive some 
negative helicity from smaller scales, but most of the transfer is from 
the forced scales ($K \approx 10$). These scales also give (negative) 
helicity to larger scales [see panels (b) and (c)]. Note that because 
helicity is negative in the large scales, a positive values of 
${\mathcal T}_h(K,Q)$ means that the shell $K$ receives negative 
magnetic helicity from the shell $Q$, increasing the absolute value 
of the magnetic helicity in the shell $K$, and the other way around 
if ${\mathcal T}_h(K,Q)$ is negative.

The forced scale [see $Q=10$ in panel (d)],
as described in the beginning of this section, is giving negative 
magnetic helicity to large scales and positive magnetic helicity 
to the small scales. This is the largest in amplitude transfer, and is the 
main source of ``absolute" magnetic helicity. 

At scales smaller than the energy injection band [see panel (e)], 
like in the magnetically forced case, (positive) magnetic helicity 
appears to cascade to smaller scales where it is finally dissipated.


\subsection{Small scale saturated regime}

\begin{figure}
\includegraphics[width=8cm]{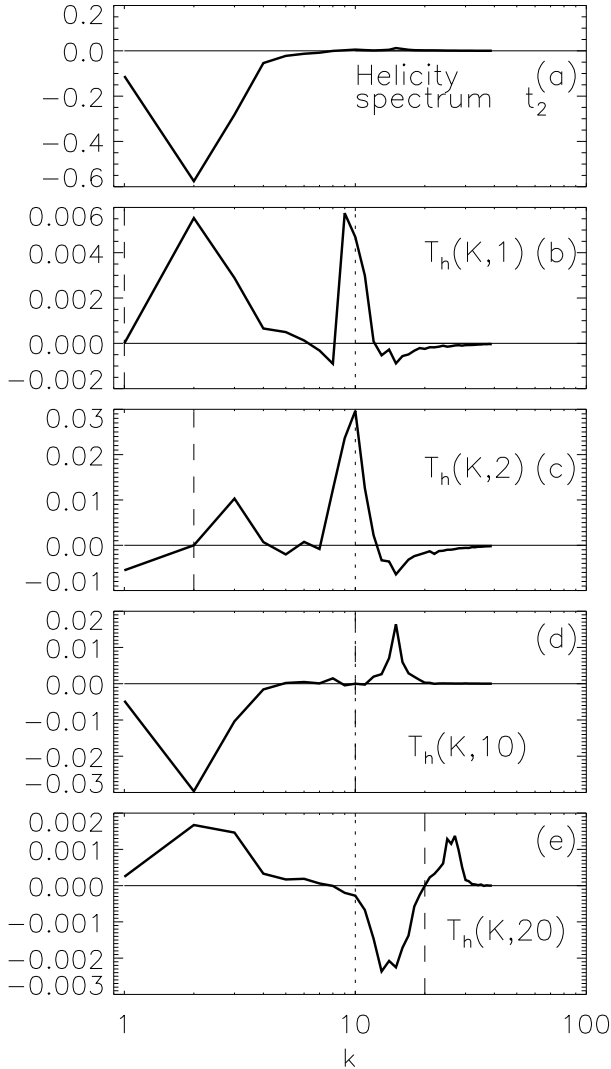}
\caption{\label{fig_06} Magnetic helicity spectrum (panel a) and 
transfer of magnetic helicity ${\mathcal T}_h(K,Q)$ normalized by the 
total magnetic helicity in the shell $Q$, for the mechanically 
forced run, when the small scales in system are saturated. The different 
panels (b-e) correspond to different values of $Q=1,2,10,20$ The dashed 
vertical line indicates the location of the examined value of $Q$, while 
the dotted line indicates the shell where the system was forced. }
\end{figure}

As the amplitude of the magnetic field is increased by dynamo action, 
the growth of magnetic energy at scales smaller than the forcing 
band saturates. Meanwhile, the negative peak of the magnetic helicity 
moves to larger scales [see Fig. \ref{fig_06}(a)]. The large scales in 
the system ($K>10$) receive (negative) magnetic helicity both locally 
from slightly smaller scales and non-locally from the forced scales,
and give negative magnetic helicity to slightly larger scales if available 
[see panel (b,c)]. The forced scale [corresponding to $Q=10$, 
see panel (d)] gives most of the negative magnetic helicity to 
the shell where the magnetic helicity spectrum peaks 
($K \approx 2$). At the same time, the forced shell gives positive 
magnetic helicity to slightly smaller scales. Finally, the small 
scales [panel (e)] cascade the positive magnetic helicity to 
even smaller scales where it is dissipated. In addition, there is a 
considerable amount of magnetic helicity destruction by 
transferring positive helicity from the small scales (where the 
magnetic helicity is mostly positive) directly into the large 
scales (where helicity is negative, see the positive peak at 
$K\approx 2$ in panel (e)), decreasing as a result the absolute 
value of magnetic helicity in both scales. We believe this 
behavior may be related to reconnection events. 

\subsection{Saturated regime}

\begin{figure}
\includegraphics[width=8cm]{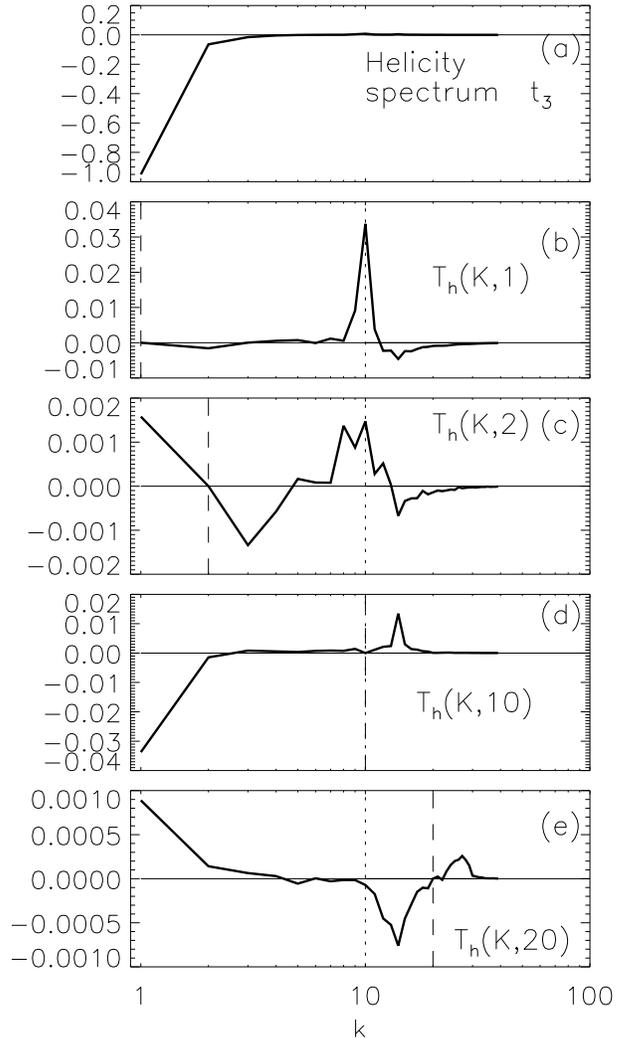}
\caption{\label{fig_07} Magnetic helicity spectrum (a), and transfer of 
magnetic helicity ${\mathcal T}_h(K,Q)$ (b-e) normalized by the total
magnetic helicity in the shell $Q$, for the mechanically forced run 
close to the saturation at the large scales. The different panels (b-e) 
correspond to different values of $Q=1,2,10,20$. The dashed vertical line 
indicates the location of the examined value of $Q$, while the dotted line 
indicates the shell where the system was forced.}
\end{figure}

When the system is close to the saturation at all scales, the 
helicity spectrum peaks at the largest available scale [$K=1$, see Fig. 
\ref{fig_07}(a)]. At this stage the large scales receive magnetic 
helicity directly from the forced scales by a non-local process [see panel 
(b) and (d)]. Such a behavior has also been observed for the transfer of 
magnetic energy in helical dynamo runs in \cite{Brandenburg01}.

In the intermediate scales, between the largest available scale in the 
box and the forced scale [see Fig. \ref{fig_07}(c)], there seems to be 
a direct cascade of helicity from the large scales to smaller scales. 
This direct cascade of helicity at large scales is similar to 
the ``reflection'' of magnetic helicity at $K=1$ observed in the 
magnetically forced run (see Sect. \ref{Magnetically_Forced}), and 
is also also expected to be dependent on the boundary conditions.

The forcing band keeps injecting magnetic helicity of opposite 
signs at large and small scales [panel (d)], but while positive magnetic 
helicity is injected at wavenumbers slightly larger than the forcing shell 
$Q=10$, most of the negative magnetic helicity is injected non-locally 
into the shell $K=1$. Scales smaller than the forced scale [see 
panel (e)] cascade the positive magnetic helicity to smaller scales 
where it is dissipated. Again, there is a non local transfer of positive 
helicity from the small scales to the largest scale (see the positive 
peak at $K=1$) leading to the decrease in the absolute value of
the magnetic helicity in both scales. Note that as the result of 
the inverse cascade of one sign of magnetic helicity 
at large scales, and the direct cascade of magnetic 
helicity of the opposite sign at small scales, 
the system is finally dominated by magnetic helicity of sign 
opposite to the kinetic helicity injected by the mechanical forcing.
This has been  observed by Brandenburg 
\cite{Brandenburg01} and in closures \cite{Pouquet76}.

\section{ \label{Concs} Discussion and Conclusions}

The results presented above stemming from the two numerical simulations have some important implications 
that need to be discussed. We start by giving a brief summary of what 
is observed in the simulations. At the early stages of the evolution of the 
magnetic field, in both examined runs the peak of the magnetic 
helicity spectrum appears to be close to the forcing scales although in 
scales slightly larger. Magnetic helicity inversely cascades in the 
large scales, both locally by transfer of helicity from the closest 
neighbor shells, and non-locally by direct transfer from the forced shells. 
As the systems evolves, the inverse cascade of magnetic helicity leads 
the magnetic helicity spectrum to peak at the largest available scale 
in the domain. At this stage, the direct input coming from the non-local 
transfer of magnetic helicity from the forced scales to the largest 
attainable scales becomes dominant. At the same time, the local transfer 
of helicity at intermediate scales changes direction, and 
magnetic helicity cascades locally to small scales. This direct 
cascade between the largest scale in the box and the forcing band can be 
expected to be sensitive to the boundary conditions, and is a 
non-universal feature  common to other systems displaying inverse 
cascade. Similar behavior has been observed in two dimensional hydrodynamic 
turbulence \cite{Smith94,Borue94,Boffetta00,Danilov01}. However, we note 
that the non-local transfer from the forced scales to the large scales is 
much greater in amplitude than the local direct cascade. This behavior 
raises the interesting question of which process, the local or non-local 
cascade, is dominant in open systems like stars or galaxies where no 
largest available scale can be clearly defined, or where stellar 
or galactic winds can eject part of the magnetic helicity out of the 
system.

The small scales behave differently. Unlike the large scales, in the 
small scales there is a noticeable direct cascade of magnetic helicity to the 
dissipation scale. This implies that in the limit of infinite Reynolds 
number in a helically forced flow, there is still going to be at the 
saturated stage a finite global magnetic helicity, since one sign of magnetic 
helicity at scales larger than the forced scales will cascade 
inversely, while the opposite sign of magnetic helicity at 
small scales will cascade to smaller and smaller scales until it will 
be dissipated.

It can be argued that this direct cascade of small scale magnetic helicity
is counter intuitive (in the sense of self-similarity), since at a 
given scale the flow does not know if it is at scales smaller than the 
forcing or larger. It could have been expected therefore to see the same 
direction of cascade at all scales. This kind of argument however assumes 
that each scale is completely independent, but this is not the case for 
MHD. Magnetic helicity in scales larger than the integral scale of a
helical flow is generated by the twisting and folding of flux tubes, 
forcing them to inter-penetrate \cite{Zeldovich}. At the same time, the 
twisting causes in small scales the magnetic field lines to spiral around 
each other, generating small scale magnetic helicity of the opposite 
sign than that in the large scale. Any further stretching of the flux tube will 
cause the small scale magnetic helicity (i.e. the twisting of the 
field lines around the flux tube) to cascade to even smaller scales, even 
if the large scale helicity is cascading to larger scales. Furthermore, 
reconnection at small scales changes the topology and the linkage of the 
field lines at the large scales, and this explains the non-local transfer 
of helicity from small scales to the large scales, ``destroying" in that 
way the large scale helicity as it is observed in Figs. \ref{fig_06} and 
\ref{fig_07}(e).

We conclude by noting that the overall picture of the cascade of 
magnetic helicity appears to be more complicated than that of the energy, 
and crucially depends on the scale and the domain size. Simple assumptions 
carried over from hydrodynamic turbulence phenomenology do not seem to 
apply here. Future numerical simulations, experiments and refined 
theoretical arguments are needed in order to illuminate further 
the understanding of MHD turbulence 
and improve the modeling of turbulent flows. 

\begin{acknowledgments}
Computer time was provided by NCAR. The NSF grant CMG-0327888
at NCAR supported this work in part and is gratefully acknowledged.
\end{acknowledgments}

\bibliography{ms}

\end{document}